# A window into the soul: Biosensing in public

Elaine Sedenberg, Richmond Wong, John Chuang

*Abstract*: Biosensed information represents an emerging class of data with the potential for massive, systematic, and remote or casual collection of personal information about people. Biosensors capture physiological signals in addition to kinesthetic data to draw intimate inferences about individuals' mental states. The proliferation of sensors makes detection, interpretation, and inference of these previously subtle – or otherwise invisible – emotional and physiological signals possible from proximate and remote locations. These sensors pose unprecedented challenges to individual privacy in public through remote, precise, and passively collected data. This paper examines the unique nature and inferential potential of biosensed data by creating a taxonomy of signals that may be collected remotely, via casual contact, or from traces left behind, and considers how these data may be collected and used to create novel privacy concerns – particularly in public. Since biosignals may often be deduced from visual data, this paper uses historic and recent photography cases to explore how social norms evolved in response to remote collection in public. A contextual integrity privacy heuristic is then used to probe the need for new norms and remedies specifically for biosensing privacy threats. This analysis examines the extensibility of relevant legal frameworks in the European Union (EU) and United States (US) as a privacy remedy, and conclude with a brief outline of possible legal or social remedies that may address privacy needs in public with biosensing technologies.

## Introduction

There is a growing public consciousness of the information trails and digital traces generated throughout our everyday lives – keystrokes, smartphones, electronic records, and interactions with digital interfaces leave millions of breadcrumbs outlining our physical paths and generate opportunities for social inferences (see, e.g., Madden and Rainie 2015). But what about our physiological and emotional states? Our heart rates, body temperatures, expressions, tones, involuntary reflexes, and biological signatures make us human yet expose our innermost nuanced physical and mental states. Though these signals have existed throughout human time, this potential for sensors to continuously record *en masse* and to systematically analyze our innate biological nature uncovers a new sensitive data class: biosensed information.

Many scholars have begun grappling with privacy and the ethical implications stemming from the quantified-self movement and the wearable industry's collection of these data (see, e.g., Crawford, Lingel, Karppi 2015). Though biosensing is commonly associated with wearables or sensors embedded into "smart" objects, little attention has been paid to the potential of these biosignals to be captured publicly from a group of non-consenting individuals over extended timescales – either remotely, through casual contact, or from biological traces left behind. Consumers of wearables have agency to put on or take off a wearable they own, while biosensing in public – outside of the consumer framework, thus bypassing consent, at least in part – frustrates our current privacy norms. Even though many of the same issues – integrity, accidental disclosure, inferential potential, generation of new knowledge, etc. – are shared across both device owners and general data subjects, the lack of consent, leaky information flows, and potential for large-scale collection and analysis by a wide spectrum of actors makes this a particularly thorny issue.

This paper examines the unique nature and inferential potential of biosensed data more broadly, to understand the urgency to consider how these data may be collected and used to create new privacy concerns, particularly in public. It then maps a subset of current biosensing technologies that may collect data remotely, via casual contact, or from traces left behind, to outline the current capabilities of biosensing technologies and root potential privacy harms within current technical capabilities. Since static and video images may be used to deduce many biosignals remotely, this paper analyzes the past evolution of social norms to mitigate or accept new privacy threats posed by developments in camera technology. Understanding the role of past social norms in mitigating harms posed by portable, consumer-grade camera devices helps illuminate the potential role norms may play within a biosensing context. Specifically given this shift in information potential without a shift in underlying technology, this paper uses Nissenbaum's (2010) "contextual integrity" privacy heuristic to fully explore the need for new norms and remedies for biosensing in a commercial public context. This analysis exposes the need to reassess the conceptual privacy risk models beyond informational privacy. Based on this examination of social norms, remedies, and exploration of biosensing's privacy risk models, this paper examines relevant legal frameworks in the European Union (EU) and United States (US) to establish the extensibility and potential gaps when applied to remote biosensing privacy harms. Finally, this paper concludes with a brief outline of possible legal or social remedies that may address privacy needs in public with biosensing technologies.

Certainly, many of the technologies and capabilities are just now on the cusp of development (past laboratory prototypes or proof-of-concept analytics): this paper calls attention to and anticipates privacy concerns relating to these emerging technologies and systems, examining what is not quite yet reality, but is far beyond mere speculation.

**The nature of biosensed data**

Biosignals represent an emerging class of data with the potential for massive, systematic, and remote or casual collection of intimate information. The proliferation of sensors makes detection, interpretation, and inference of these previously subtle – or otherwise invisible – emotional and physiological signals possible from proximate and remote locations. Although these cues are essential to biological survival and have existed for millennia, technology stands to sense and collect these data on orders of magnitude more perceptive than the human eye, and on extended timescales impossible through casual observation.

Biosensing technologies are being developed to capture a wide range of physiological signals from which inferences may be drawn, including our mental and emotional states, predispositions to diseases, or proclivities to particular behaviors. Many of these states are unknown or subconscious to us, and many of these patterns are not yet fully understood by modern science. The applications and business models emerging from biosensed information will raise new security and privacy challenges that are not yet fully comprehended or anticipated.

This new emerging class of data sets itself apart from existing information types through a set of unique characteristics.

- **Expansive in scope**. Measurable signals are diverse and include physiological (e.g., electroencephalogram (EEG), electrocardiogram (ECG), heart rate (HR), electromyography (EMG), electrodermal activity (EDA, otherwise known as Galvanic Skin Response or GSR), functional near-infrared spectroscopy (fNIRS), blood pressure, blood glucose) and kinesthetic (e.g., accelerometry, eye-tracking, facial expression recognition). From combinations of these signals, inferences about the body or actions can be drawn to deduce emotional states, physical reactions, and even memories and thoughts.
- **Intimate yet leakable**. Biosensors may be wearables (e.g., commercial fitness trackers worn on the wrist, which require opt-in and are removable) as well as remote (e.g., heart rate monitoring through surveillance video feeds, which do not require contact or consent and which may be invisible to the data subject), and they may be standalone or embedded in everyday objects. Given biosignals may be naturally – and largely uncontrollably – emitted as signatures (e.g., electrical, chemical, or biological) or as involuntary impulses, the information flows are leaky in the sense that diverse actors (or multiple actors at once) may sense them remotely or through brief contact.
- **Precise yet ambiguous**. Biosensors are increasingly able to produce readings with many significant figures, yet the high-level inferences drawn from these raw signals will be context dependent and highly ambiguous. For instance, a single emotion may have different associations or spectrums for one individual compared to others.
- **Familiar yet unverifiable**. When shown a photo, audio clip, or geolocation tag of themselves, it is usually fairly straightforward to confirm these data match the ground truth of a data subject's experience. However, if shown a biosignal, such as an EEG recording, it would be much more difficult to confirm its provenance relating to themselves, not to mention verify its accuracy. Further, experiences may not always map to the measurements: one may feel a racing heart at the sight of a romantic crush, but it may or may not correspond to an increase in heart rate.
- **Limited controllability**. Many of these signals, like those associated with respiration, or involuntary reflexes are controlled by the autonomic nervous system and are difficult to consciously control. Our biological signals, like fingerprints or genetic profiles, are similarly difficult to control since individuals leave traces through casual contact. Some of these signals, impulses, and signatures may be obscured through conscious overriding (e.g., deep breathing) or obfuscated by physical barriers.

Unlike many other information flows that require computer-mediated interaction, biosensed data are emitted as (or may be inferred from changes in) thermal and electromagnetic radiation, visual or observable cues, biochemical signals and traces, or encoded in biological signatures. While sensors may be voluntarily worn by consumers or self-recorded and collated by quantified self-enthusiasts, developing and emerging technologies enable even more data streams to be collected remotely, through casual contact – meaning sensors that may be briefly, and perhaps unknowingly, touched through everyday interaction – or through physical traces left behind. This potential makes biosensed data collection with no opportunity for consent a new privacy challenge for individuals in public spaces.

For each potential type of biosensed information, sensors or diagnostics can be operated from a distance to the individual data subject, after the subject has left a trace, or through casual contact that might be mediated by everyday interactions like touching a door handle, or sitting on a bench. This paper specifically aims to focus on the potential of biosensing technology apart from wearable devices in order to fully grapple with information flows that may involve no prior consent or knowledge from the individual data subject. Remote biosensing may be deployed on top of existing sensor technology – such as cameras – by introducing new analytical processes that uncover previously undetected biosignals, or through the deployment of entirely new sensor technology.

*Table 1 presents a taxonomy of sensors and potential capabilities by type of signal, measurable biosensed phenomena, inferential potential, and relevance to remote, traceable, or casual contact public collection.*

Table 1: A taxonomy of sensors and potential capabilities

| Measurable phenomena or characteristic | Methods of measurement | Example potential inferences | Distance and interaction potential |
|---|---|---|---|
| Body temperature | Infrared thermal radiation | Activity level; health status; acute changes in stress levels | Remote via infrared cameras or infrared thermometers (spot ratio for accurate measurement is several feet away) (Scigiene Corporation, n.d.). |
| Heart rate | Electromagnetic radiation; radio frequency (electromagnetic fields) | Activity level; health status; changes in excitement or stress | Vital Radio (several feet away, possible through walls) (Adib et al. 2015; Mullan et al. 2015; Wang et al. 2015). Remote and smartphone cameras also may be able to detect pulse and heart abnormalities (Hewitt 2013; Li et al. 2014) (several feet away; eyeshot) |
| Eye tracking | Infrared and near-infrared non-collimated light reflections | Attention; gaze; drowsiness; traumatic brain injuries | Some eye tracking technology requires a stable head, but new devices do not, or take advantage of other devices on the head (e.g., glasses, headset). Measurement could be taken from a high-resolution camera a few feet in front of the subject. |
| Retinal scan | Infrared light absorption | Identity; presence of health conditions (e.g., diabetes, stroke, hypertension, heart disease, neurovascular disease, etc.) (MacGillivray et al. 2014). | Close proximity and bright light. |
| Iris scan/pupil | Camera with | Identity; arousal; health | High-resolution images of the eye |
| dilation | infrared illumination or electromagnetic radiation | status (e.g., stroke history, or in children leukocoia, an early sign of retinoblastoma) (Clusmann, Schaller, Schramm 2001; Palca 2014). | – potentially through proximate but still remote means via other technologies like Virtual Reality (VR) headsets. |
| Facial recognition | Electromagnetic radiation | Identity; presence of genetic disorders (Cooper 2014). | Low-resolution images of a face. |
| Emotion | Electromagnetic radiation | Mental or emotional states. | Video or images of face may be made from a distance to capture microexpressions. |
| Fingerprints | Image patterns for identifying characteristics | Identity | Traces may be left behind during casual contact, but fingerprints may be sensed through contactless scanning in one second (Biotime Technology 2014). |

Each measureable phenomena on its own – such as heart rate, temperature, pupil dilation, etc. – may not contain much inferential power since many of these signals will rise and fall in concert to obvious external stimuli. For example, several measures may increase simultaneously if an individual becomes startled. Insights may be inferred more easily, however, when systems change asynchronously. If heart rate speeds up but another measure like blood oxygenation or EDA stays constant, these data may reflect a mental over an autonomic change of state. Additionally, any change in an individual's usual baseline may indicate a notable change. Biometric authentication that conducts facial recognition or fingerprints – which is already done on some banking apps – would also inherently come with the potential to give private companies access to other biosensed information (Corkery 2016). Given the nature of biosensed data, voluntarily giving access to one feature like an iris scan may enable countless other inferences to be made simultaneously.

The inferential potential, beyond mere physiological responses derived from these sensors and analytics, is only beginning to come to light. Leaning upon small-scale studies – the current state of the art in this area of research – one can understand how these biosensing systems may be able to uncover unknown emotional or physiological states hidden in plain view. Our microexpressions lie underneath outward facades or in fleeting looks, and betray concealed or subconscious emotional states that are too short or subtle to notice with the naked eye (Adler 2015; Li et al. 2015). Outside of biosensing, there are proof-of-concept academic studies that demonstrate an ability to predict health conditions based on language use alone. One research group used language from Twitter (in particular, the use of angry language) to predict heart disease mortality at a community level. Another study conducted a case study analysis of former US Presidents Ronald Reagan and George H. Bush to retroactively compare for differences in discourse complexity that may be related to early Alzheimer's disease (Berisha et al. 2015). It is notable that inferences like these are able to retroactively examine existing corpuses to deduce future health states. With biosensing technology, individuals may produce larger, more intimate data repositories that reveal not only more emotional granularity than a public tweet, but will last into the future, making retroactive analyses possible. Young adults may build biosensed dossiers in their twenties that may be relevant in 30 years for unforeseen analytics. For instance, researchers studied the predictive associations of resting heart rate on violent and non-violent criminality using the measurements of over 700,000 Swedish men when they began mandatory military conscription at age 18. Researchers found a low resting heart rate was associated with an increased risk for violent criminality, non-violent criminality, risk of being a victim of an assault, and being involved in accidents as an adult (Latvala et al. 2015). The full potential for insights and the range of possible uses remain unknown, but the intimacy and raw potential are certain.

**Sensing and social norms**

Often, the introduction of new technologies grates against pre-existing social norms, and societies are given a relatively short timeframe to react and adjust laws accordingly (Tene and Polonetsky 2013). Bicchieri (2000) describes two separate interpretations of social norms: in one norms are descriptive of what is considered normal behavior and, in the other, norms are injunctive of what the public believes should be done for social approval or disproval. Laws may develop to propagate social norms and further prevent social harm – as was the case in seat belt laws – and laws may also develop to create torts or other possible remedies for socially unacceptable harms (Licht 2008).

Biosensed information, particularly remote biosensing to collect data without any physical contact, frustrates many of the norms for recording in public places without individuals' knowledge or consent. This is especially true given that biosensing is entirely new (at least outside of the medical context) and is experiencing rapid technological and analytical development. The following subsections explore two prior cases of remote sensing in public – specifically, the capture of information using photography beginning with the amateur and portable Eastman Kodak Camera, and followed by the Google Glass embedded camera. Historically images could capture expressions and situations, or tie individuals to particular locations – now images (static or video) may unlock new inferences and measurements of various biosignals. Each photography case provides insight into potential social norms that may arise with remote biosensing technologies, and begins to explore how these norms collide and influence laws and policies.

*Introduction of the Kodak Camera*

The introduction of the Eastman Kodak Camera in 1884 made photography portable and affordable to the general public, and coincided with the rise of sensationalistic journalism and large print circulations (Solove and Schwartz 2015). Photography was brought into homes and family life, and Kodak advertising promoted the technology in the household as a signal of modernity and leisure (Sarvas and Frohlich 2001, 55). However, this new technology opened private spaces and fleeting moments up to permanent scrutiny, in some ways to certify our memories or impressions of others (West 2000, xi). In response to public fears and normative tensions regarding the potential invasion of cameras and relative permanency of images, resort beaches and even the Washington Monument banned their use temporarily until law and social norms caught up (Lindsay 2000).

Published in 1890, the canonical article "The Right to Privacy" by Warren and Brandeis explicitly states: "Thoughts, emotions, and sensations demanded legal recognition," and calls attention to recent inventions and business methods, "instantaneous photographs and newspaper enterprise," as threats to the sacred precincts of private and domestic life (Warren and Brandeis 1890, 195). This article established the need for a new right for privacy, which was not protected under common law but could be modified using torts to redress harms.

The introduction of the camera clearly illustrates a time where social norms were forced to rapidly adapt to the presence of a new ability to record information (images that relayed our facial expressions, locations, social interactions, and actions) about ourselves and others. Similarly, biosensing technology presents a new information to record physiological and emotional data in previously public, invisible, or fleeting spaces – often using the same portable image technology or similarly small, and amateur friendly recording devices. Existing legal frameworks were not prepared to protect and remedy situations brought on by the collision of law with technological advances, and adaptations had to be made in order to keep up with evolving social norms and values around privacy. At present, it seems absurd – if not impossible – to ban photography in public spaces like the Washington Monument. Yet there are still social situations where it would be seen as socially inappropriate to take photos of others, even it is technically legal.

*Google Glass*

Google Glass, first publicly announced in April 2012, and released to some members of the public in a public beta program in April 2013, provides insight into user and media perspectives on emerging technologies and the privacy concerns people may raise in anticipation of the release of new devices. After the announcement of Glass in 2012, people imagined what it might be like to live in a world with Glass, a world reflected in Glass's portrayal in Google's promotional videos and advertisements and in broader media discourse. Through the process of speculating and anticipating the future with Glass, people began raising concerns about remote sensing and privacy (Wong and Mulligan 2016).

Google's initial conceptual video of Glass was released on April 4, 2012 on YouTube.com. The video is shot entirely in a first person point of view, only showing the viewpoint of a single male Glass user as he makes his way around New York City. It never shows what the actual device looks like, rendering the device invisible. Furthermore, the user is never seen putting Glass on or taking it off, suggesting that the device is worn and turned on all the time, and not taken off or turned off between interactions with it. The video shows the user walking in the street, at a bookstore, and meeting a friend at a food truck. Notably, these are all in public or semi-public spaces. Together, this video begins to portray Glass as a device that seems to fade into the background, is always worn and always turned on throughout the day, capable of recording pictures and video at any time, and used across many contexts by one person. Against this backdrop, people began discussing concerns about privacy in several ways. First, people expressed concerns about surveillance and surreptitious recording by (other) Glass users, noting that people could be secretly taking pictures of others in varied public places including the subway, the street, airplanes, or at beaches, clubs, and bars (Hill 2012). These concerns were facilitated by the perception of Glass as an invisible, seamless, and mobile technology. Second, people expressed concerns about how data collected by Google through Glass might be used. Some, contemplating

Google's business model, suggested that Glass would be used to collect personal data to deliver more advertisements (Thier 2013) or that Glass would share data with other companies like LinkedIn or Facebook to, for example, conduct facial recognition in public (Burns 2013). A third set of concerns was raised about Glass breaching structural social and technical constraints that protect privacy (Surden 2007). Perceiving Glass as mobile, always-on, and blending into the background suggested that prior structural constraints – such as social norms, limited camera battery life, and the visibility of smartphones – that prevented certain types of data collection might change with Glass. Ephemeral public experiences in everyday life might unwillingly and unknowingly become digitally archived. A fourth set of concerns about Glass transgressing social contexts was raised. Glass's mobility and always-on portrayal suggested that it was untethered to context, leading to fears that information could be taken from one context and easily used in another without regard for the social norms that usually govern information flows in a given context.

As time got closer to the launch of the Google Glass beta program, some people began taking private remedies to protect their privacy, including private business owners banning Glass in bars and strip clubs (DiSalvo 2013), and the formation of an online campaign trying to ban Glass called Stop the Cyborgs.

In many ways, biosensing technology is at a similar stage of development – still under development but with public exposure. Like with Glass, the people whose biosignals are being monitored and collected may not be able to see the sensors and may not even know when or where this collection is occurring. Biosensing much like Glass also raises questions about data flows – what people, companies, and third parties are able to access the data, and how might that data be used. Biosensing also begins to erode some structural constraints that protected privacy – signals that were previously protected because no technologies or techniques existed to record them can now be recorded. Moreover, networked biosensing technologies have the potential to violate contextual norms. Perhaps, the Glass example also suggests that when new technologies and technology products are announced and publicized, productive public conversations can occur about the social implications of those new technologies.

**Remote biosensing in context**

After discussing various social norms and responses that are associated with protecting privacy and exploring historical and recent cases with similarities to biosensing technologies, this paper now turns to examine emerging aspects of remote biosensing that may be problematic. Nissenbaum's contextual integrity heuristic provides a useful framework to compare the changes in data flows afforded by remote biosensing through new analytical power on existing technology and introduction of novel sensors. Contextual integrity holds that rather than framing privacy as control and prevention of information or data flows, privacy is better understood as a set of relations and information flows governed by "context-relative informational norms" (Nissenbaum 2010). That is, different social contexts have varying social norms that govern appropriate flows of information and data – and the violation of these norms suggests the potential violation of privacy. Four key steps emerge as a heuristic from this framework: 1) establish the prevailing social context; 2) establish the key actors involved in the information and data transfer (including data senders, data recipients, and data subjects); 3) indicate what attributes, or data types, are involved; 4) establish the principles of data transmission. If new practices introduce changes to any of these elements, it suggests a potential violation of contextual integrity. This provides a way to understand and assess how *changes* in practices may affect informational norms.

The contextual integrity heuristic is used here to investigate a plausible deployment of remote biosensing technologies at a shopping mall in the US. While the precise context described in this scenario depicts a situation in a near, proximate future, it is grounded in real, developing technologies, testbeds, prototypes, and pilot projects. By exploring these possibilities while technologies and systems are still in development, concerns related to privacy and discussions of potential remedies can be raised in a proactive, rather than reactive manner, at a time when systems may still be amenable to change. This follows traditions across many fields, most notably in recent calls for "privacy by design," which advocates for addressing privacy concerns through the engineering, design, and use of technologies through technical, business, and legal means ( Cavoukian 2011; Computing Community Consortium 2015; Mulligan and King 2011; Rubinstein 2011).

*Social context*

Imagine a number of remote biosensors from Table 5.1 being used at a shopping mall somewhere in the US. This includes infrared cameras that can detect people's body temperatures, cameras and wireless devices that can sense heart rate, and video cameras that can scan faces for pupil dilation, facial recognition, or emotion. In this example, a user- or customer-centric point of view is taken, focusing on practices that include the sensing of customers. Uses of these remote biosensing technologies might include presenting marketing and advertisements to customers, monitoring for security, collecting data for analysis, one-party actions (such as one user sensing a second user's signals without their knowledge or consent; or a customer service representative sensing a customer), or two-party interpersonal actions (such as two users sensing and sharing their signals with each other). While shopping malls are privately owned, shopping malls are still semi-public spaces, outside the sites of the private domestic home or private workplace.[1] Prior research on human–computer interaction and computer mediated social interaction in public space tends to take a broader view of public spaces, including semi-public sites such as cafes, bars, train stations, and restaurants, which serve the same social functions of public space of sociality and recreation (Humphreys 2005; Humphreys 2010). From a user or consumer point of view, these sites constitute being "in public." Thus following contextual integrity's focus on the perspective of users, this example similarly considers a broader view of "public space" to include semi-public sites of social interaction, such as shopping malls.

*Actors*

The data subjects in this context are the mall customers. Their data is sensed and collected, although they may or may not consent to the collection of their data (or may consent without full knowledge of the capabilities of technical sensors, inferences, or full knowledge of legal protections they may or may not have). The data collectors vary, but include the mall operators, mall security, store workers and, potentially, other customers that have access to these sensors. Data recipients include all the data collectors, but also include the companies that process and return the data collected by sensors, and potential third party services and companies that store, access, or further process the data.

---

[1]    See, e.g., *Pruneyard Shopping Center v. Robins*, 447 U.S. 74 (1980).

*Attributes*

In a mall today, without remote biosensors, several attributes – or data types – are already in play. Photos and videos of customers are recorded for surveillance purposes. Customers' real-time locations can be collected and aggregated using cell phone data (Clifford and Hardy 2013). During financial transactions, customers may share personal information such as their names, addresses, and credit card numbers. With remote biosensed information, additional attributes are invoked as well – both those collected directly by the sensors and those that may be inferred from that data. These sensors may also collect photographic or video images, but at resolutions that allow pupil measurement and the inference of one's identity, emotions, arousal, and health status. Body temperature and infrared images can be collected as well, which may be used to analyze activity level and health status. Heartbeat and breathing activity data can also be collected, providing insight into activity level, health status, excitement, and stress.

*Principles of transmission*

The contextual integrity heuristic asks us to contemplate current baseline norms that provide the principles of transmission and compare those to any changes in the principles of transmission. The principles of transmission are affected by current practices, laws, technological capabilities, and social norms. Drawing on Solove's taxonomy of privacy harms, the processes of "transmission" can encompass multiple stages of the data lifecycle, including data collection, processing, and dissemination (Solove 2006). The analysis of transmission principles in this example draws upon prior research studying user behaviors with technology and attitudes toward data collection and sharing in public (and semi-public) spaces, current practices, current technical capabilities of widely implemented systems, and laws and regulations that signify and help set social norms. While acknowledging that there are a range of behaviors around how individuals perceive and perform their identities and emotions, this analysis focuses on some of the present principles surrounding the stage of collection, then around the stages of data processing and dissemination. It then explores how the collection, processing, and dissemination of biosensed data may change those particular principles.

Currently, surveillance of physical spaces is primarily visual and camera-based (Räty 2010). Furthermore, in research studying users' perceptions of data collection with new augmented reality devices in public, users make comparisons to already existing visual recording technologies, including cellphone cameras and CCTV cameras (Denning, Dehlawi, and Kohno 2014). This suggests that practices and expectations around these existing visual technologies can be used to understand baseline norms that provide principles of transmission. Currently, security cameras can only capture what is within their line of sight, and the cameras' locations are generally visible to customers. Due to the cameras' visibility by virtue of their physical placement as well as, frequently, the presence of posted notices, customers can choose to avoid an area or avoid a camera's line of sight should they not want to be recorded, or customers may elect to change their behavior or self-presentation (Massimi et al. 2010). Video surveillance is also conducted at a resolution that might allow facial recognition – either computationally or by other humans, though with limited accuracy and consistency (Trottier 2014; Introna and Nissenbaum 2009) – but not at a resolution that would allow for iris detection or pupil measurement. These technical limitations of video surveillance create structural protections for privacy, that is, privacy is implicitly protected because certain technical capabilities do not yet exist or impose heavy costs to implement (Surden 2007). These limitations also create normative expectations about what is possible and acceptable to monitor and collect.

In interpersonal interactions inside the mall, people expect that they can remain pseudonymous and not reveal their identity while walking around; in other words, people expect information about their identity to *not* be collected. People expect that they can interact interpersonally without revealing their identity as well, such as asking an employee for help or asking another person for directions. Individuals have control over when to reveal their identity. In these interactions, people's "body language" might be interpreted, but it cannot be measured or independently verified. One's ground truth emotions and mental states are generally considered to be personal and generally vague in nature – even if sometimes outwardly expressed or revealed in fleeting reactions. Prior work has found that people act in accordance to social norms that allow others to have private conversations in public spaces. For instance, Humphreys (2005) finds that often in public spaces, when someone is conducting a conversation on a mobile phone, he or she will turn away or shield their face while other people in their party will occupy themselves with other activities – even if they are able hear the conversation – in order to acknowledge the social norms of private conversation in public. Social norms also limit what types of recording activities are deemed appropriate in public spaces – for instance, nonconsensual, perverse video recordings of others in public spaces are deemed unacceptable socially. In many states in the US, voyeurism laws support these social norms concerning inappropriate uses of cameras (Solove and Schwartz 2015).

Transmission principles regarding the processing and dissemination of data collected in a public space such as a shopping mall include data sharing with third parties only in particular circumstances and maintaining data security when the data are shared. While many people expect to be recorded by CCTV cameras in public spaces (Nguyen, Kobsa, and Hayes 2008), people's expectations suggest that collected data should not be widely shared for processing and dissemination. Users tend to expect that the people with access to recordings of public spaces include security personnel, employees, owners, and the police, but do not expect the material to be distributed to other actors (Massimi et al. 2010). Other studies corroborate this, finding that participants view camera installations in public spaces to be less acceptable if footage is streamed to a remote location (Friedman et al. 2006). While surveillance camera data may be shared with law enforcement (with a warrant), traditional CCTV surveillance camera footage is stored locally.

Today, internet connected (IP-based) security cameras already challenge these transmission principles, as the video data could easily be stored offsite with third parties. These types of distributed architectures also allow for new types of processing, such as automated or crowdsourced facial recognition (Introna and Nissenbaum 2009; Trottier 2014). These emerging forms of data processing may create changes to the transmission principles, by allowing new actors access to the data, violating subjects' reasonable expectations, or changing how privacy risks and benefits are distributed among people and groups (Introna and Nissenbaum 2009).

When interacting with retailers, customers expect their personal financial information – such as credit card number, name, and address – to be processed by external institutions, but this is done in the context of a transaction, such as a purchase. Engaging in this transaction provides implicit consent for the transmission and use of this information for purposes of the transaction. Customers may also expect the store to collect purchase data to understand customers' habits, possibly using a third party to process the information. The results of this information processing may be used as feedback to try to emotionally manipulate customers' purchasing habits through pricing strategies, "sales" signs, physically arranging displays in particular ways, or using music or aromas in the store (Klosowski 2013). However, this data processing (and resulting feedback interventions) are done at aggregate levels, either on all customers or on broad demographics such as age or gender. Data subjects also generally expect that data security precautions are taken during the processing of their data. Principles of data security are also established through technical frameworks and standards that companies follow, such as the ISO 27001 standard (ISO 2017) or NIST Cybersecurity Framework (NIST 2014).

The introduction of remote biosensing technologies into the mall setting may disrupt these principles and norms. Biosignals are *intimate yet leakable*, allowing remote biosensors to collect the data without sustained contact and without a line of vision, such as Vital Radio, which wirelessly detects heartbeats. Sensors that are not cameras may not look like sensors either. Without these visual cues, customers may not be aware that the sensors even exist or that their signals are being collected, processed, and disseminated – by the mall, by stores and their employees, or by other individual customers. The uses of biosensors described above may allow stores to build biosensed profiles over time about individual customers and target advertisements or products directly to individuals.

The *limited controllability* of biosignals suggests that information that was once latently protected and hidden, such as identity, emotions, body temperature, activity level, or stress, are now open to being collected by a wide range of actors including mall management, store employees, and other customers. This information can be collected when one is simply in the presence of the biosensors, outside of the contexts of financial transactions or interpersonal interactions. Thus, it becomes easy to capture biosignals without customers' knowledge or consent, whether explicit or implicit. This means that customers may lose their ability to walk around a mall anonymously and they may lose plausible deniability about their emotional states.

The actors who receive the information for processing and dissemination are likely to expand. Because biosensed information is *precise yet ambiguous*, the data will likely be processed by and disseminated to third parties, such as data analytics companies or by the companies that create the technologies. These data are then interpreted and returned back to the end user. In many cases, the end users will not be the data subjects. For instance, a store may use a set of biosensors to collect data on its customers, and then use third-parties to analyze those data. This creates a new long-term and ongoing relationship between end users and the data recipients who process and disseminate the data. This stands in contrast to the traditional model where the company who sold a surveillance camera to a mall did not get full access to the recordings of every camera it sold. Furthermore, mall customers are unlikely to know of these data processing recipients or that their data is being sent to and analyzed by these recipients. The technical capabilities or new biosensors make it likely that this type of data processing, dissemination, and feedback can occur continuously rather than being tied to particular transactions, and can occur at an individual level rather than an aggregate one. Together, these shifts in the transmission principles suggest new types of privacy harms that may occur with the deployment of biosensing technologies in public spaces.

**Biosensing: beyond information privacy**

As highlighted above, in the examples of early photography and Google Glass, public fears about the ability to capture unsavory emotions or surreptitiously record individuals is particularly poignant in the case of biosensing technology. By using contextual integrity as a heuristic, the previous section further explored how remote biosensing in a public context may challenge and frustrate social and legal norms. Based upon the public context for biosensing, this paper now considers how biosensing may require a new risk model for privacy extending beyond information privacy alone.

Drawing upon Tavani's (2008) categorization of four distinct types of privacy – (1) physical/accessibility, (2) decisional, (3) psychological/mental, and (4) informational – this section briefly examines the potential for biosensing in public to infringe on more than one meaning of privacy. Physical privacy, which focuses primarily on privacy as physical non-intrusion, may be considered the least relevant to public biosensing. Even in the cases where casual and unforced contact might be needed to mediate a sensor reading, there would be no mandatory or disruptive physical impact to the person.

Decisional privacy may be considered the freedom from interference affecting choices like education, health care, career, personal life, and beliefs (Floridi 2006). Biosensed information easily gives outsiders clues to individuals' reactions, attitudes towards an object or situation, or belies one's ambivalence about a situation. Gathering biosensed data creates an opportunity to interfere with or manipulate results. For instance, in Mexico, political campaigns have used biosignals like heart rate, facial coding, and neuro-feedback (collected from EEG data in the lab since, as of 2018, EEG data cannot be remotely or casually sensed) to assess voter attitudes and reactions to particular campaign material (Randall 2015). Collecting these data remotely in the wild to tailor advertising, including campaign material, is not far off. Smart-shelf technology has also been developed by companies like snack maker Mondelez International Inc. to collect real-time demographic and behavioral data about customers in order to present a customized ad experience that makes impulsive junk-food purchases even more tempting (Boulton 2013; Truth Labs, 2014). These decisional interventions could be made to manipulate and interfere with countless areas of life, or be designed to prey upon our behavioral vulnerabilities like addictions.

The psychological or mental privacy to protect our intimate thoughts remains fairly straightforward for biosensed data. Scientists and evolutionary psychologists have studied why some cues are made public to assist us in public settings. For instance, tears are thought to be a subtle cue to instigate empathy from others within a few feet, but when biosensed data like emotions or internal physiological states are systematically recorded and analyzed, all signals become magnified beyond their original natural public scope. Further, inferences drawn from data individuals may not necessarily know or feel within themselves not only violates their inner most thoughts but, importantly, may circumvent our internal mental states altogether by presenting an algorithmic determination back to us, thus preventing our own emotional processes. The ways in which biosensing data in public could both intrude and interfere with our mental states and cause harm are numerous and present new challenges.

The informational meaning of privacy remains the most similar to other modern challenges toward limiting access and allowing control over one's personal information. Biosensing creates the potential for a greater amount of an entirely new data class of intimate information to be gleaned at a high rate from public encounters and kept for indiscriminate amounts of time. The ways in which information privacy may result in inappropriate flows of information and potential harms are numerous and vast.

Though Warren and Brandeis (1890) bring up emotions specifically in the call for a right to privacy, current legal and policy frameworks may not be able to fully cover privacy harms beyond information privacy. The ability to ascertain and make judgments (undoubtedly subjective and with bias) about internal emotional states using biosensing technology and using these inferences to interfere with decisional autonomy creates additional legal tensions relating to privacy.

**Sensing and legal frameworks**

The fuzziness around what type of data biosensing information may be categorized as, emerging inferential power, along with the rapidly expanding potential for new remote technologies (that may be both inexpensive, discrete, and able to capture new unfamiliar data types) frustrate current legal frameworks. This paper focuses on EU and US laws due to their respective prominence in global privacy law and contrasting approaches.

The forthcoming EU General Data Protection Regulation (GDPR),[2] which will supersede the Data Protection Directive[3] in May 2018, will likely prevent almost all public collection and processing of individuals' biosignals without explicit permission by the individual. "Personal data" is defined as "any information relating to an identified or identifiable natural person" where an identifiable person is one who can be directly or indirectly identified by referencing identifiers like name, ID number, location, or using one or more factors specific to that person – including physiological, genetic, mental, and social identity.[4] The regulation specifically calls out "biometric data" as a special category of personal data subject to heightened scrutiny.[5] Biometric data is defined by the regulation as "personal data resulting from the specific technical processing relating to physical, physiological, or behavioral characteristics of a natural person, which allow or confirm the unique identification of that natural person, such as facial images or dactyloscopic data."[6]

---

[2]     Regulation (EU) 2016/679, 2016 O.J. (L 119) 1.
[3]     Directive 95/46/EC, 1995 O.J. (L 281) 31 (EU).
[4]     Regulation (EU) 2016/679, art. 4, para. 1, 2016 O.J. (L 119) 1, 33.
[5]     Regulation (EU) 2016/679, art. 9, 2016 O.J. (L 119) 1, 38.
[6]     Regulation (EU) 2016/679, art. 4, para. 14, 2016 O.J. (L 119) 1, 33.

These definitions could be problematic because not all biosensing data may be identifiable or able to authenticate an individual but could still be revealing – a heightened heartrate signal can still be used to help make real-time assessments about a response or state even if it is not uniquely identifiable or combined with personal information. Though the ability to isolate biosignals from other indirect identifiers like location data is unlikely.

Further, it is possible for biosensing to frustrate EU restrictions on the collection of biosignals information by satisfying any number of exceptions listed under Article 9 of the GDPR. The law allows an exception if "processing relates to personal data which are manifestly made public by the data subject."[7] As discussed in this paper, many of these signals are of limited control and outward facing, and thus may be considered "public" within the context of this law. Explicit consent from the data subject allows for the processing of personal data, and thus biosignals that are tied to identity. Given the familiar yet ambiguous and inferential potential of these signals, it is not unlikely that someone may consent to having their heart rate recorded without understanding how it may be used to make inferences about their health or emotional states. Consent also may be given without complete understanding of the ambiguous nature of these signals and the potential for improper categorization, like elevated heartrate inferring poor health state instead of genetic variability. Even with the ability to easily revoke consent, subjects may not fully understand the implications or inherent ambiguity of these signals in a way that motivates action to do so. The personal and "household use" of these data would also be permitted, which further underscores the importance of social norms in negotiating privacy in these unregulated social contexts.

Biosignals could also challenge broader clauses within the law, like processing that is "necessary for the purposes of preventative or occupational medicine," or for medical assessment. Many biosignals could lead to important biomedical inferences about individuals, and government, public health officials, or private healthcare providers may be able to use these justifications as means to wider collection and use. There may be additional exemptions such as those related to national security that open up other uses of biosignals. Even with the most comprehensive privacy laws to date, public biosignal collection may still pose privacy harms within the EU and spur further debates regarding the definition and classification of biosignals as a new class of sensitive data. The numerous ways in which GDPR may fully cover or leave opportunities for biosignal use is beyond the scope of this paper, but will be important in future research as the policy and biosensing technologies mature.

Unlike the EU, the US has no single data protection law but has legislated on an *ad hoc* basis for particular sectors or circumstances. Privacy coverage under the Health Insurance Portability and Accountability Act (HIPAA)[8] would only be potentially extended to biosignals collected by organizations subject to the "privacy rule" – i.e., those defined as covered entities.[9] In situations where a patient presents their own collected biosignals to a healthcare professional, it is unclear at what point, if any, those data may be considered covered information. In general in the US, biosignals associated with wearable "wellness" devices are considered low risk and the Food and Drug Administration (FDA) does not regulate the production and use of these devices. It could be possible that a law similar to HIPAA could regulate the exchange and storage of biosensing data when held by particular entities, but the coverage of potential data collectors and holders would be limited.

---

[7]     Regulation (EU) 2016/679, art. 9, para. 2(e), 2016 O.J. (L 119) 1, 38.

[8]     Health Insurance Portability and Accountability Act of 1996, Pub. L. No. 104–191, 1996 U.S.C.C.A.N. (110 Stat.) 1936 (codified at 42 U.S.C. §§ 1320d-1 to d-9 (2012)).

[9]     45 C.F.R. §§ 160, *et seq.*, 164.104 (2014).

In the US, it could be possible for a privacy law like the Video Voyeurism Prevention Act of 2004[10] to prohibit the particular use of a type of technology. In response to the rise of cell phone cameras that were discrete enough to allow surreptitious image capture, notably around loose garments of unconsenting individuals, states and the federal government responded with video voyeurism laws to make this particular use of legal cameras illegal. A legal restriction on particular information gathering activities using otherwise legal instruments could theoretically be possible to prevent particular data collection from remote biosensing technologies.

Finally, within the US, the Federal Trade Commission (FTC) would likely be the strongest enforcer of privacy infringements through its mandate to prevent unfair and deceptive commercial practices (FTC 1980; FTC 1983). Cases brought by the FTC would be responsive to the undisclosed use, misuse, or insecure storage of biosensed data by commercial actors in the US. While cases may be reactive to harms caused to consumers, it may be possible for the FTC to issue guidance on responsible uses of biosensing technology and data similar to how the agency has done recently in regards to the internet of things (FTC 2015).

**Conclusion: a consideration of remedies**

In some ways, the ability to remotely collect emotional profiles or record others in public is similar to traditional photography and the tensions highlighted in the Google Glass debate. However, the unique nature of biosensed data as an emerging data class with particular attributes regarding meaning, integrity, and availability complicate existing social norms and privacy laws. As further demonstrated using contextual integrity, the ability to record remotely in public ushers in new tensions and highlights gaps in current legal structures that may not fully consider the ability to infringe upon decisional and emotional privacy in addition to expected information privacy concerns.
The possible remedies related to biosensing in public require additional research and debate. As discussed in this paper, existing privacy laws and regulations in the EU and US contain gaps that could allow for harm by biosensing technologies and the collection, use, and sharing of associated data. However, broad restrictions on the use of personal information, like those in the EU, could help stem some harms linked to making inferences about an individual's mental or physical states, and additional regulation on the use of these technologies in non-identifying situations may be necessary. Better privacy protection could be afforded by restricting particular collection activities or biosensing technology uses, similar to the laws regulating video voyeurism with cameras. These laws could potentially act as a deterrent or create the opportunity for remedies of harms done. Additionally, protecting the storage and exchange of biosensing data in a HIPAA style law could prevent some harms by providing standards and protections for these data, but may be limited in the range of entities covered. Retroactive regulation for harms conducted by private industry or perhaps individuals will certainly be important as biosensing becomes more pervasive, but will necessarily be reactive instead of preventative.

---

[10]     18 U.S.C. § 1801 (2004).

Other potential remedies beyond law and policy may be important to adapting social norms. Personal modulation as a form of individual obfuscation may be possible and serve as a remedy against remote intrusion. Instrument jammers, data spoofers, or physical blocks may be worn to prevent some types of biosensing measurements. Masks may be used to prevent emotional analysis, and photography may be muddled by the use of reflective scarves that block images taken with flash (Access All Brands 2017). Though these are not geared specifically for biosensing, they represent opportunities individuals may be able to take in the absence of legal or social control. These remedies, and the roles they may have in developing social norms, should be explored in future work.

Community standards may be an option to complement social norms as a soft form of technology regulation. Social community standards may make particular actions using biosensing technology inappropriate and socially unacceptable due to peer pressure or public ridicule – or vice versa making evolving biosensing applications acceptable. Additionally, there may be an opportunity to develop technical standards into biosensing technology that allow a value to become embedded into the design – similar to efforts to make Do Not Track a standard for webpages online. This would likely be considered more of a technical standard setting effort as opposed to social community standard, but could offer a similar remedy.

Anticipating new privacy concerns in public spaces due to the rise in remote biosensing technologies, our use of the contextual integrity heuristic identifies several potential privacy concerns, and discussed potential remedies. This is just a first step in a broader conversation about privacy, ethics, and values – in the future, various ethical considerations may determine that other interests and values outweigh the privacy concerns of these technologies. Yet these conversations must occur actively (and proactively), among technologists, policymakers, and the general public so that technical, legal, and social decisions are made explicitly.

**Acknowledgements**

This work was funded and made possible by a generous grant from the Hewlett Foundation and Center for Long-Term Cybersecurity (CLTC) at the University of California, Berkeley. The authors would also like to thank the Berkeley BioSENSE Lab Group for their comments and feedback in the early stages of this manuscript proposal, the reviewers and book editors for their thoughtful and constructive feedback, and research assistant Matt Nagamine for his support and attention to detail.
This material is based upon work supported by the National Science Foundation Graduate Research Fellowship Program under Grant No. DGE1106400. Any opinions, findings, and conclusions or recommendations expressed in this material are those of the author(s) and do not necessarily reflect the views of the National Science Foundation.

# References

Access All Brands. 2017. "ISHU: Anti-Flash Scarf." Accessed January 5. https://theishu.com/.

Adib, Fadel, Hongzi Mao, Zachary Kabelac, Dina Katabi, and Robert C. Miller. 2015. "Smart Homes that Monitor Breathing and Heart Rate." *Proceedings of the 33rd Annual ACM Conference on Human Factors in Computing Systems – CHI '15*, 837–846.

Adler, Jerry. 2015. "Smile, Frown, Grimace and Grin – Your Facial Expression Is the Next Frontier in Big Data." *Smithsonian.com*, December. Accessed January 5, 2017. http://www.smithsonianmag.com/innovation/rana-el-kaliouby-ingenuity-awards-technology-180957204/

Berisha, Visar, Shuai Wang, Amy LaCross, and Julie Liss. 2015. "Tracking Discourse Complexity Preceding Alzheimer's Disease Diagnosis: A Case Study Comparing the Press Conferences of Presidents Ronald Reagan and George Herbert Walker Bush." *Journal of Alzheimer's Disease 45*(*3*): 959–963.

Bicchieri, Cristina. 2000. "Words and Deeds: A Focus Theory of Norms." In *Rationality, Rules, and Structure*, edited by Julian Nida-Rümelin and Wolfgang Spohn, 153–184. Dordrecht: Kluwer Academic Publishers.

Biotime Technology. 2014. "Finger on the Fly: Contactless 4 Fingerprint Capture in Less than 1 Second Now Available." *YouTube.com*, December 14. Accessed January 5, 2017. https://www.youtube.com/watch?v=lhAe36m4jQs

Boulton, Clint. 2013. "Snackmaker Modernizes the Impulse Buy with Sensors, Analytics." *CIO Journal* (blog), *Wall Street Journal*, October 11. https://blogs.wsj.com/cio/2013/10/11/snackmaker-modernizes-the-impulse-buy-with-sensors-analytics/

Bradfield, Ron, George Wright, George Burt, George Cairns, and Kees Van Der Heijden. 2005. "The Origins and Evolution of Scenario Techniques in Long Range Business Planning." *Futures 37*(*8*): 795–812.

Burns, Will. 2013. "Nine Business Applications for Google Glass." *Forbes*, February 27. http://www.forbes.com/sites/willburns/2013/02/27/nine-business-applications-for-google-glass-love-ideasicle/

Cavoukian, Ann. 2011. "Privacy by Design: The 7 Foundational Principles." Accessed November 1, 2017. https://www.ipc.on.ca/wp-content/uploads/Resources/7foundationalprinciples.pdf

Clifford, Stephanie, and Quentin Hardy. 2013. "Attention, Shoppers: Store Is Tracking Your Cell." *New York Times*, July 14. http://www.nytimes.com/2013/07/15/business/attention-shopper-stores-are-tracking-your-cell.html

Clusmann, H., C. Schaller, and J. Schramm. 2001. "Fixed and Dilated Pupils after Trauma, Stroke, and Previous Intracranial Surgery: Management and Outcome." *Journal of Neurology, Neurosurgery & Psychiatry 71*(*2*): 175–181.

Computing Community Consortium. 2015. "Visioning Activity: Privacy by Design." Accessed January 6, 2017. http://cra.org/ccc/visioning/visioning-activities/2015-activities/privacy-by-design/

Cooper, Charlie. 2014. "Facial Recognition Technology Used to Spot Genetic Disorders." *Independent*, June 23. http://www.independent.co.uk/news/science/facial-recognition-technology-used-to-spot-genetic-disorders-9558032.html.

Corkery, Michael. 2016. "Goodbye, Password. Banks Opt to Scan Fingers and Faces Instead." *New York Times*, June 21. http://www.nytimes.com/2016/06/22/business/dealbook/goodbye-password-banks-opt-to-scan-fingers-and-faces-instead.html

Crawford, Kate, Jessa Lingel, and Tero Karppi. 2015. "Our Metrics, Ourselves: A Hundred Years of Self-Tracking from the Weight Scale to the Wrist Wearable Device." *European Journal of Cultural Studies 18*(*4–5*): 479–496.

Denning, Tamara, Zakariya Dehlawi, and Tadayoshi Kohno. 2014. "In Situ with Bystanders of Augmented Reality Glasses: Perspectives on Recording and Privacy-mediating Technologies." *Proceedings of the 32nd Annual ACM Conference on Human Factors in Computing Systems*, 2377–2386.

DiSalvo, David. 2013. "The Banning of Google Glass Begins (And They Aren't Even Available Yet)." *Forbes*, March 10. http://onforb.es/10rxJBl

FTC (Federal Trade Commission). 1980. "FTC Policy Statement on Unfairness." Federal Trade Commission, December 17. Accessed January 5, 2017. https://www.ftc.gov/public-statements/1980/12/ftc-policy-statement-unfairness

FTC (Federal Trade Commission). 1983. "FTC Policy Statement on Deception." Federal Trade Commission, October 14. Accessed January 5, 2017. https://www.ftc.gov/system/files/documents/public_statements/410531/831014deceptionstmt.pdf

FTC (Federal Trade Commission). 2015. "Internet of things: Privacy & security in a connected world." Federal Trade Commission, January. Accessed January 5, 2017. https://www.ftc.gov/system/files/documents/reports/federal-trade-commission-staff-report-november-2013-workshop-entitled-internet-things-privacy/150127iotrpt.pdf

Floridi, Luciano. 2006. "Four Challenges for a Theory of Informational Privacy." *Ethics and Information Technology 8*(*3*): 109–119.

Food and Drug Administration. 2016. "General Wellness: Policy for Low Risk Devices – Guidance for Industry and Food and Drug Administration Staff." July 29. Accessed January 5, 2017. https://www.fda.gov/downloads/MedicalDevices/DeviceRegulationandGuidance/GuidanceDocuments/UCM429674.pdf

Friedman, Batya, Peter H. Kahn, Jr., Jennifer Hagman, Rachel L. Severson, and Brian Gill. 2006. "The Watcher and the Watched: Social Judgments about Privacy in a Public Place." *Human-Computer Interaction 21*(*2*): 235–272.

Guston, David H. 2014. "Understanding 'Anticipatory Governance.'" *Social Studies of Science 44*(*2*): 218–242.

Hewitt, John. 2013. "MIT Researchers Measure Your Pulse, Detect Heart Abnormalities with Smartphone Camera." *ExtremeTech*, June 21. Accessed January 5, 2017. http://www.extremetech.com/computing/159309-mit-researchers-measure-your-pulse-detect-heart-abnormalities-with-smartphone-camera

Hill, Kashmir. 2012. "Sergey Brin's Favorite Google Glass Feature." *Forbes*, September 11. http://onforb.es/RIXia6

Humphreys, Lee. 2005. "Cellphones in Public: Social Interactions in a Wireless Era." *New Media & Society 7*(*6*): 810–833.

Humphreys, Lee. 2010. "Mobile Social Networks and Urban Public Space." *New Media & Society 12*(*5*): 763–778.

ISO (International Organization for Standardization). 2017. "ISO/IEC 27001 Family – Information Security Management Systems." Accessed January 6, 2017. http://www.iso.org/iso/iso27001

Introna, Lucas, and Helen Nissenbaum. 2009. "Facial Recognition Technology: A Survey of Policy and Implementation Issues." The Center for Catastrophe Preparedness and Response, New York University, July 22. https://ssrn.com/abstract=1437730

Klosowski, Thorin. 2013. "How Stores Manipulate Your Senses So You Spend More Money." *Lifehacker*, April 18. Accessed January 6, 2017. http://lifehacker.com/how-stores-manipulate-your-senses-so-you-spend-more-mon-475987594

Latvala, Antti, Ralf Kuja-Halkola, Catarina Almqvist, Henrik Larsson, and Paul Lichtenstein. 2015. "A Longitudinal Study of Resting Heart Rate and Violent Criminality in more than 700000 Men." *JAMA Psychiatry 72*(*10*): 971–978.

Li, Xiaobai, Jie Chen, Guoying Zhao, and Matti Pietikäinen. 2014. "Remote Heart Rate Measurement from Face Videos under Realistic Situations." *Proceedings of the IEEE Computer Society Conference on Computer Vision and Pattern Recognition*, 4264–4271.

Li, Xiaobai, Xiaopeng Hong, Antti Moilanen, Xiaohua Huang, Tomas Pfister, Guoying Zhao, and Matti Pietikäinen. 2015. "Towards Reading Hidden Emotions: A *Comparative Study of Spontaneous Micro-Expression Spotting and Recognition Methods.*" *IEEE Transactions on Affective Computing PP* (99): 1–13.

Licht, Amir N. 2008. "Social Norms and the Law: Why Peoples Obey the Law." *Review of Law and Economics 4*(*3*): 715–750.

Lindsay, David. 2000. "The Kodak Camera Starts a Craze." PBS. Accessed January 5, 2017. http://www.pbs.org/wgbh//amex/eastman/peopleevents/pande13.html

MacGillivray, Tom, Manuel Trucco, James Cameron, Baljean Dhillon, Graeme Houston, and Edwin van Beek. 2014. "Retinal Imaging as a Source of Biomarkers for Diagnosis, Characterisation and Prognosis of Chronic Illness or Long-Term Conditions." *The British Journal of Radiology 87*: 1–16.

Madden, Mary, and Lee Rainie. 2015. "Americans' Attitudes about Privacy, Security and Surveillance." *Pew Research Center*, May 20. Accessed January 5, 2017. http://www.pewinternet.org/2015/05/20/americans-attitudes-about-privacy-security-and-surveillance/

Massimi, Michael, Khai N. Truong, David Dearman, and Gillian R. Hayes. 2010. "Understanding Recording Technologies in Everyday Life." *IEEE Pervasive Computing* 9(*3*): 64–71.

Mullan, Patrick, Christoph M. Kanzler, Benedikt Lorch, Lea Schroeder, Ludwig Winkler, Larissa Laich, Frederik Riedel, et al. 2015. "Unobtrusive Heart Rate Estimation during Physical Exercise Using Photoplethysmographic and Acceleration Data." *Proceedings of the Annual International Conference of the IEEE Engineering in Medicine and Biology Society, EMBS 2015-November*, 6114–6117.

Mulligan, Deirdre K., and Jennifer King. 2011. "Bridging the Gap between Privacy and Design." *University of Pennsylvania Journal of Constitutional Law*, 989–1034.

NIST (National Institute of Standards and Technology). 2014. *Framework for Improving Critical Infrastructure Cybersecurity*. NIST. https://www.nist.gov/sites/default/files/documents/cyberframework/cybersecurity-framework-021214.pdf

Nguyen, David H., Alfred Kobsa, and Gillian R. Hayes. 2008. "An Empirical Investigation of Concerns of Everyday Tracking and Recording Technologies." *Proceedings of the 10th international conference on Ubiquitous computing (UbiComp '08)*, 182–191.

Nissenbaum, Helen. 2010. *Privacy in Context*. Stanford: Stanford University Press.

Palca, Joe. 2014. "Chemist Turns Software Developer after Son's Cancer Diagnosis." *NPR*, May 6. http://www.npr.org/sections/health-shots/2014/05/06/309003098/chemist-turns-software-developer-after-sons-cancer-diagnosis

Randall, Kevin. 2015. "Neuropolitics, Where Campaigns Try to Read Your Mind." *New York Times*, November 3.http://www.nytimes.com/2015/11/04/world/americas/neuropolitics-where-campaigns-try-to-read-your-mind.html.

Räty, Tomi D. 2010. "Survey on Contemporary Remote Surveillance Systems for Public Safety." *IEEE Transactions on Systems, Man, and Cybernetics, Part C (Applications and Reviews) 40*(*5*): 493–515.

Rubinstein, Ira S. 2011. "Regulating Privacy by Design." *Berkeley Technology Law Journal 26*: 1409–1456.

Sarvas, Risto, and David M. Frohlich, eds. 2011. "The Kodak Path (ca. 1888–1990s)." In *From Snapshots to Social Media – The Changing Picture of Domestic Photography*, 47–82. London, UK: Springer London.

Scigiene Corporation. n.d. "Infrared Thermometers." Accessed January 5, 2017. http://www.scigiene.com/pdfs/Infrared%20Thermometers.pdf

Solove, Daniel. 2006. "A Taxonomy of Privacy." *University of Pennsylvania Law Review 154*(*3*): 447–560.

Solove, Daniel, and Paul M. Schwartz. 2015. *Information Privacy Law*. New York: Wolters Kluwer.

Surden, Harry. 2007. "Structural Rights in Privacy." *Southern Methodist University Law Review 60*: 1605–1629.

Tavani, Herman T. 2008. "Informational Privacy: Concepts, Theories, and Controversies." In *The Handbook of Information and Computer Ethics*, edited by Kenneth E. Himma and Herman T. Tavani, 131–164. Hoboken, NJ: John Wiley & Sons, Inc.

Tene, Omer, and Jules Polonetsky. 2013. "A Theory of Creepy: Technology, Privacy and Shifting Social Norms." *Yale Journal of Law and Technology 16*(*1*): 59–102.

Thier, Dave. 2013. "The Potentially World-Changing New Feature in Google's Project Glass." *Forbes*, February 20. http://onforb.es/11VjbyK

Trottier, Daniel. 2014. "Crowdsourcing CCTV Surveillance on the Internet." *Information, Communication & Society 17*(*5*): 609–626.

Truth Labs. 2014. "Mondelez Smart Shelf." *Vimeo.com*. Accessed January 5, 2017. https://vimeo.com/96148396

Wang, Siying, Antje Pohl, Timo Jaeschke, Michael Czaplik, Marcus Kony, Steffen Leonhardt, and Nils Pohl. 2015. "A Novel Ultra-Wideband 80 GHz FMCW Radar System for Contactless Monitoring of Vital Signs." *Proceedings of the Annual International Conference of the IEEE Engineering in Medicine and Biology Society*, 4978–4981.

Warren, Samuel D., and Louis D. Brandeis. 1890. "The Right to Privacy." *Harvard Law Review 4*(*5*): 193–220.

West, Nancy Martha. 2000. *Kodak and the Lens of Nostalgia*. Charlottesville: University of Virginia Press.

Wong, Richmond Y., and Deirdre K. Mulligan. 2016. "When a Product Is Still Fictional: Anticipating and Speculating Futures through Concept Videos." *Proceedings of the 2016 ACM Conference on Designing Interactive Systems*, 121–133.